\documentclass[pra,amsmath,amssymb,showpacs,superscriptaddress,twocolumn]{revtex4}
\input epsf.tex
\usepackage{graphicx}
\usepackage{amsthm}
\usepackage{array}
\usepackage{longtable}
\usepackage{dcolumn}
\usepackage{bm,bbm}
\usepackage[usenames,dvipsnames]{color}
\usepackage{amsmath}
\usepackage[hypertex]{hyperref}
\usepackage{amsfonts}
\usepackage{amssymb}
\usepackage{mathrsfs}
\usepackage{subfigure}
\usepackage{ulem}

\begin{document}

\title{Tight finite-key analysis for passive decoy-state quantum key distribution under general attacks}

\affiliation {Zhengzhou Information Science and Technology Institute, Zhengzhou, 450004, China}
\affiliation {Key Laboratory of Quantum Information,University of Science and Technology of China, Hefei, 230026, China}
\affiliation {Synergetic Innovation Center of Quantum Information and Quantum Physics, University of Science and Technology of China, Hefei, Anhui 230026, China}

\author {Chun Zhou} \author {Wan-Su Bao}\email{2010thzz@sina.com}
\affiliation {Zhengzhou Information Science and Technology Institute, Zhengzhou, 450004, China}
\affiliation {Synergetic Innovation Center of Quantum Information and Quantum Physics, University of Science and Technology of China, Hefei, Anhui 230026, China}

\author {Hong-Wei Li}
\affiliation {Zhengzhou Information Science and Technology Institute, Zhengzhou, 450004, China}
\affiliation {Key Laboratory of Quantum Information,University of Science and Technology of China, Hefei, 230026, China}
\affiliation {Synergetic Innovation Center of Quantum Information and Quantum Physics, University of Science and Technology of China, Hefei, Anhui 230026, China}

\author {Yang Wang} \author {Yuan Li}
\affiliation {Zhengzhou Information Science and Technology Institute, Zhengzhou, 450004, China}
\affiliation {Synergetic Innovation Center of Quantum Information and Quantum Physics, University of Science and Technology of China, Hefei, Anhui 230026, China}

\author {Zhen-Qiang Yin} \author {Wei Chen} \author {Zheng-Fu Han}
\affiliation {Key Laboratory of Quantum Information,University of Science and Technology of China, Hefei, 230026, China}
\affiliation {Synergetic Innovation Center of Quantum Information and Quantum Physics, University of Science and Technology of China, Hefei, Anhui 230026, China}

 \date{\today}

\begin{abstract}
For quantum key distribution (QKD) using spontaneous parametric-down-conversion sources (SPDCSs), the passive decoy-state protocol has been proved to be efficiently close to the theoretical limit of an infinite decoy-state protocol. In this paper, we apply a tight finite-key analysis for the passive decoy-state QKD using SPDCSs. Combining the security bound based on the uncertainty principle with the passive decoy-state protocol, a concise and stringent formula for calculating the key generation rate for QKD using SPDCSs is presented. The simulation shows that the secure distance under our formula can reach up to 182 km when the number of sifted data is $10^{10}$. Our results also indicate that, under the same deviation of statistical fluctuation due to finite-size effects, the passive decoy-state QKD with SPDCSs can perform as well as the active decoy-state QKD with a weak coherent source.
\end{abstract}

\maketitle

\section{introduction}
Quantum key distribution (QKD) allows two legal communication parties to acquire the identical key based on quantum mechanics. Since the invention of the first pioneer QKD protocol, BB84 protocol \cite{BB84}, people have achieved great progress both in QKD's theory and experiment \cite{Lo99,GLLP04,Tomamichel11,Guo12,Zeilinger12,Bacco13,Pan13,Shield13}. On the way toward the industrialization of QKD, people have faced sorts of obstacles, one of which comes from the fact that the necessary assumptions required for QKD's unconditional security are not easy to satisfy in a real situation \cite{Scarani09}. Practical factors, i.e., inefficient authentication of classical communication, imperfections of setups and finite-size data, will undoubtedly threaten the security of a real QKD system and quantum hacking strategies can be successfully derived to attack the practical QKD system \cite{Hacking1,Hacking2,Hacking3}. However, corresponding countermeasures can be applied to combat these attacks. One approach is to employ the notion of device-independent QKD (DI-QKD) \cite{DI1,DI2,DI3,MDI1,MDI2,MDI-Lo,MDI-Ma,Bao13-1}. The other, although difficult to implement, is to mathematically characterize the impact of imperfect factors on QKD's security as comprehensively as possible by a security proof\cite{Scarani09}.

The notion of finite-length keys is one of practical imperfections need to be solved in the practical security of QKD. In the case of finite-length keys, the security bound in the asymptotic regime should be reconsidered, and several attempts have been made to tackle this problem \cite{Hayashi07,Cai09,Li09,Song11,Somma13}. In recent years, based on the composable security definition derived from trace distance \cite{Renner05}, several significant advances have been achieved \cite{Scarani08-1,Scarani08-2,Christandl09,Bratzik11,Ng12,Hayashi12,Hayashi13}, with the most pioneering one being the bound from the smooth min-entropy by Scarani and Renner \cite{Scarani08-2}. By noting that the uncertainty relation can be generalized to one formulated in terms of smooth entropies and that this directly implies the security of QKD protocols \cite{Tomamichel11}, Tomamichel et al. \cite{Tomamichel12} creatively introduced the entropic formulation of the uncertainty relation into the security analysis of finite-length keys. Since then, many attempts were made to improve the security bound of finite resources, such as the situations for permutation-invariant protocols under coherent attacks \cite{Mertz13}, active decoy-state QKD \cite{Lim13}, measurement-device-independent QKD \cite{Curty13}, one-sided device-independent QKD \cite{Bao13-2}, and the B92 protocol \cite{Mafu13}. It should be noted that, by applying generalized chain rules for smooth min-entropies \cite{Vitanov13}, information leakage from multiphoton pulses and vacuum pulses that the eavesdropper may exploit can be well bounded \cite{Lim13}. Thus, the result of Ref. \cite{Tomamichel12} can be applied to most real situations when practical photon sources are used, e.g., weak coherent sources (WCSs) and spontaneous-parametric-down-conversion sources (SPDCSs).

SPDCS, like the commonly used WCSs, is also within reach of current technology and can be considered as another candidate of the perfect single photon source. However, due to the multiphoton fraction, QKD using SPDCSs is also vulnerable to the photon-number-splitting attack \cite{Brassard00}. The active decoy-state method \cite{decoy03,decoy-Lo,decoy-Wang}, i.e., actively and randomly varying the intensity of each signal state by a variable optical attenuator (VOA), can be conducted to combat this attack. But in some cases the imperfections of VOA might cause some physical parameters to rely on the particular setting selected and then threaten QKD's security \cite{Curty10}. Thus, passive preparation of intensity might be desirable in practice, and the first passive decoy-state protocol was presented by introducing a photon number resolving detector \cite{Pas-Mauerer07}. Then, Adachi et al. \cite{Pas-Adachi07} presented an efficient passive decoy-state proposal (AYKI protocol) which can be easily realized with a practical threshold detector. More importantly, it is proved to be efficient enough for estimating the contribution of the single-photon pulse. Later, Ma and Lo \cite{Pas-Ma08} generalized the results of Refs. \cite{Pas-Adachi07} and \cite{Pas-Mauerer07} to the most common case and Curty et al. \cite{Pas-Curty09} proposed a new passive decoy-state scheme for QKD using WCSs by subtly fitting a beam splitter with a threshold detector for triggering. However, all of the above results regarding the passive decoy-state scheme are obtained in the condition of asymptotic infinite-length keys. An effort to derive the security bound for the passive decoy-state method under finite resources has been made by Tan and Cai \cite{Tan11}. Their work is based on an indirect approach of tracing coherent attacks to collective attacks by the de Finetti theorem \cite{Renner07}. And for the direct approach based on the uncertainty principle \cite{Tomamichel12}, how the finite-size effect influences the performance of a passive decoy-state protocol needs further studying. This is just what we intend to clarify here.

In this paper, we directly introduce the formula of key generation rate obtained from Ref. \cite{Lim13} into the cases for passive decoy-state protocol. The difference is in the parameter estimation step, i.e., the way to estimate the single-photon yield and error rate in the scenario of finite-length keys. The starting point of the passive decoy-state protocol under asymptotic infinite-length keys is that the yield and bit error rate of the $n$-photon states from the triggered pulses are both equal to that from the nontriggered pulses. But this condition is no longer true under the condition of finite-length keys due to statistical fluctuations. Hence, we shall reconsider the steps of a passive decoy-state protocol for estimating single-photon yield and error rate. Luckily, it is found that the yield and bit error rate of $n$-photon states can be considered as random variables emanating from sampling without replacement. Then, by applying the Serfling bound \cite{Serfling74} in sample theory, one can construct confidence regions of the interval estimate for these variables, which was first introduced into the parameter estimation of QKD by Scarani et al. \cite{Scarani08-2} and then improved by Tomamichel et al. \cite{Tomamichel12} and Mertz et al. \cite{Mertz13}. Thus, in the confidence regions, there certainly exist relationships for the parameters between triggered events and nontriggered events, which can be directly applied to estimate the gain and bit error rate of triggered and non-triggered single-photon events, respectively. In particular, without relying on any approximation, we introduce a rigorous method based on a hypergeometric argument \cite{Hayashi12} to bound the quantity of the maximal information of an eavesdropper on the single-photon events. Note that our security analysis is conducted based on the uncertainty principle and that bound in \cite{Hayashi12} holds true under no approximation; thus the formulas we obtain are valid for general coherent attacks and our results guarantee unconditional security. We compare our results with those derived from active decoy protocol \cite{Lim13} and the simulations show the efficiency of our protocol.

The paper is organized as follows. In Sec. II, we fix the security preliminaries, clarify the formalism used to calculate secret key rates under the assumption of general attacks and introduce the bound for estimating the phase error rate in our protocol. Section III recalls the AYKI protocol for QKD with asymptotic infinite-length keys. The main results of this paper, i.e., tight formulas for estimating the yield and bit error rate for single-photon events, are presented in Sec. IV. Section V numerically simulates our results and Sec. VI concludes the paper.

\section{Security criteria and smooth min-entropy}
In this paper we consider an asymmetric coding BB84 protocol, where the bases $X$ and $Z$ are chosen with probabilities $q_X$ and $q_Z$ that are biased. The protocol consists of these steps: state preparation, state measurement, sifting, parameter estimation (PE), error correction (EC), error verification, and privacy amplification (PA) (for a detailed description, see Ref. \cite{Lim13}). The protocol outputs are $S_A$ and $S_B$ on Alice's and Bob's side respectively. Only and only if successfully passing all of the above steps can $S_A$ and $S_B$ be considered secure. Here, the security criterion based on trace distance, seminally proposed by Renner, is introduced in our analysis \cite{Renner05}:

\textit{Definition 1 (composable security definition)}. Assume a QKD protocol outputs keys of $S_A$ and $S_B$ on Alice¡¯s and Bob¡¯s side respectively. It is considered to be $\varepsilon-\text{secure}$ if it satisfies both the correctness and the secrecy.
Correctness means that the protocol is $\varepsilon_{cor}$-correct if $P(S_A \neq S_B)\le \varepsilon_{\text{cor}}$, namely the probability of $S_A\neq S_B$ will not exceed $\varepsilon_{\text{cor}}$.
Secrecy means that the protocol is $\varepsilon_{\text{sec}}-\text{secret}$ if $\frac{p_{\text{pass}}}{2}{{\left\|{{\rho}_{\text{SE}}}-U_{S}\otimes {{\rho }_{E}} \right\|}_{1}}\le \varepsilon_{\text{sec}}$ where $S$ represents either of the keys $S_A$ and $S_B$, $\rho_E$ is the system that the eavesdropper owns, ${\rho}_{SE}$ is the classical-quantum state describing the joint state of $S$ and $E$, $U_S$ is the uniform mixture of all possible values of $S$, and $p_{\text{pass}}$ is the probability that all steps of the protocol are successfully conducted.

Smooth min-entropy, relying on a generalization of the von Neumann entropy, is an essential tool in the security proof based on information theory \cite{Renner05}. Combined with the uncertainty principle, it directly implies a security proof without the assumption that the measurement devices work according to the specifications of the protocol \cite{Tomamichel11}. In particular, it can provide us an efficient method for the finite-key analysis \cite{Tomamichel12}. If we denote $\mathcal{H}$ as a finite dimensional Hilbert space and let $\mathcal{P}(\mathcal{H})$ be the set of positive semidefinite operators on $\mathcal{H}$. Then, the set of normalized quantum states and subnormalized ones can be presented by $\mathcal{S}(\mathcal{H}):=\{\rho\in\mathcal{P}(\mathcal{H}):\text{tr}\rho=1\}$ and $\mathcal{S}_{\leqslant}(\mathcal{H}):=\{\rho\in\mathcal{P}(\mathcal{H}): \text{tr}\rho\leqslant1\}$, respectively. Given theses, the definition of smooth min-entropy can be defined as the following\cite{Renner05}:

\textit{Definition 2 (smooth min-entropy)}. Let ${\varepsilon}\ge 0$, $\sigma_{B}\in\mathcal{S}(\mathcal{H}_{B})$ and $\rho_{AB}\in\mathcal{S}_{\leqslant}(\mathcal{H}_{AB})$. The smooth min-entropy $H_{\min}^{{\varepsilon}}(A|B)$, taken over a set of states $\mathcal{B}^{\varepsilon}(\rho)$ that are $\varepsilon$-close to $\rho_{AB}$, is defined as the quantity
\begin{equation}
\begin{aligned}
{\max_{\tilde{\rho}\in{\mathcal{B}^{\varepsilon}(\rho_{AB})}}}\{-\log_2\min\{\lambda>0:\exists \sigma_B :\tilde{\rho}_{AB}\le \lambda \text{id}_{A}\otimes \sigma_{B}\}\},
\end{aligned}
\end{equation}
where $\mathcal{B}^{\varepsilon}(\rho):=\{\tilde{\rho}_{AB}\in\mathcal{S}_{\leqslant}(\mathcal{H}_{AB}):C(\rho_{AB},\tilde{\rho}_{AB})\leqslant \varepsilon\}$, $\text{id}_{A}$ is the identity operator on $A$, $C(\rho_{AB},\tilde{\rho}_{AB}):={1-(\text{tr}|\sqrt{\rho}\sqrt{\tilde{\rho}}|)^2}^{1/2}$ is a distance measure based on fidelity and ${\varepsilon }$ is called the smoothing parameter.

There exists the following chain rule for the smooth min-entropy\cite{Vitanov13,Dupuis14}:

\textit{Lemma 1 (Chain-rule inequality for the smooth min-entropy)}. Let ${\varepsilon}\ge 0$, ${\varepsilon'}, {\varepsilon''}\ge 0$, and $\rho_{ABC}\in\mathcal{S}_{\leqslant}(\mathcal{H}_{ABC})$. Then
\begin{equation}
\begin{array}{lll}
H^{\varepsilon+\varepsilon'+2\varepsilon''}_{\min}(AB|C)_{\rho} \geq H^{\varepsilon''}_{\min}(A|BC)_{\rho}+ H^{\varepsilon'}_{\min}(B|C)_{\rho}\\
\quad\quad\quad\quad\quad\quad\quad\quad\quad\quad-f(\varepsilon),
\end{array}
\end{equation}
where $f(\varepsilon)=\log_2{\frac{1}{1-\sqrt{1-\varepsilon^2}}}$.

Let system $E'$ be the information that Eve obtains on the raw key $X_A$ of Alice, prior to the error-verification step. Then, after the privacy amplification step, the length of the secure key $S_A$ can be expressed by the following lemma.

\textit{Lemma 2 (Secret key based on smooth min-entropy) \cite{Renner05,Lim13}:} By applying privacy amplification with two-universal hashing, a secret key extracted from $X_A$ is $\varepsilon_{\text{sec}}$-secret if its length $\ell$ is chosen such that
\begin{equation}
\begin{array}{lll}
\lfloor H^{\nu}_{\min}(X_A|E')-2\log_2{\frac{1}{2\overline{\nu}}} \rfloor,
\end{array}
\end{equation}
where $\nu+\overline{\nu}\leq \varepsilon_{\text{sec}}$ with $\nu$ and $\overline{\nu}$ chosen to be proportional to $\frac{\varepsilon_{\text{sec}}}{p_{\text{pass}}}$, and $H^{\nu}_{\min}(X_A|E')$ quantifies the amount of uncertainty system $E¡ä$ has on $X_A$.

\section{AYKI protocol with infinite-length keys}
In the AYKI protocol, two-mode states are prerequisite and we consider thoses emitted from the nondegenerate spontaneous-parametric-down-conversion (SPDC) process. This type of SPDC processes creates the two-mode state \cite{SPDC00}
\begin{equation}
\begin{array}{lll}
(\cosh\chi)^{-1}\sum\limits_{n=0}^{\infty }{{{(\tanh \chi )}^{n}}{{e}^{in\theta }}\left| n,n \right\rangle }.
\end{array}
\end{equation}
Set the intensity $\mu$ of the source to $\sinh^2\chi$, then the above description simplifies to
\begin{equation}
\begin{array}{lll}
\sum\limits_{n=0}^{\infty}{\sqrt{\frac{{{\mu}^{n}}}{{{(1+\mu )}^{n+1}}}}{{e}^{in\theta }}\left|n,n\right\rangle}
\end{array}
\end{equation}

When the sender (Alice) measures one mode of her states from the above SPDCS with a practical threshold detector described by detection efficiency $\eta_A$ and dark-count rate $d_A$, the other mode can be divided into two parts according to the response of the threshold detector, i.e., the triggered events and nontriggered events. Both of the them are sent to the lossy channel, detected by the receiver's (Bob's) detector and devoted to the final secret key. In particular, the nontriggered events, acting as the role of decoy states, can be used to estimate the single-photon contribution and single-photon error.

In this case, the signal $n$-photon events with probability $p_n$ are also divided into two parts, the triggered $n$-photon events with probability of $p^{(\text{t})}_n$ and the nontriggered $n$-photon events with probability of $p^{(\text{nt})}_n$. Let $\gamma_n$ be the probability of detection (triggering) when $n$ photons are emitted from the SPDC process. Then, $p^{(\text{t})}_n=p_n\gamma_n$ and $p^{(\text{nt})}_n=p_n(1-\gamma_n)$ with \cite{Pas-Adachi07}
\begin{equation}
\begin{array}{lll}
p_n=\frac{\mu^n}{(1+\mu)^{n+1}},\quad \gamma_n=1-(1-d_A)(1-\eta_A)^n.
\end{array}
\end{equation}

In this paper, we consider the measurement model mentioned in Ref. \cite{Pas-Adachi07}. It should be noted that, in the case of asymptotic infinite-length keys, it is assumed that the detection rate (yield) and quantum bit error rate (QBER) of the triggered $n$-photon events are the same as those of the nontriggered $n$-photon events, i.e.,
\begin{equation}
\begin{array}{lll}
Y^{(t)}_n=Y^{\text{(nt)}}_n, \quad\quad e^{(t)}_n=e^{\text{(nt)}}_n.
\end{array}
\end{equation}
Under this condition, it is not easy to find that \cite{Pas-Adachi07}
\begin{equation}
\begin{array}{lll}
Q^{(t)}_n=\delta_nQ^{\text{(nt)}}_n,
\end{array}
\end{equation}
where $Q^{(t)}_n=Y^{(t)}_n p_n \gamma_n$, $Q^{\text{(nt)}}_n=Y^{\text{(nt)}}_n p_n (1-\gamma_n)$ and $\delta_n=\frac{\gamma_n}{1-\gamma_n}$. Noting that $0\leq \delta_0 <\delta_1 <\delta_2 <\cdots$ and considering the overall detection rate $Q^{(t)}=\sum^{\infty}_{n=0}Q^{(t)}_n$ with triggering and $Q^{\text{(nt)}}=\sum^{\infty}_{n=0}Q^{\text{(nt)}}_n$ without triggering, one can obtain a lower bound for the single-photon detection rate $Q^{\text{(nt)}}_1$ without triggering \cite{Pas-Adachi07}:
\begin{equation}
\begin{array}{lll}
Q^{\text{(nt)}}_1\geq\frac{(\delta_2-\delta)Q^{\text{(nt)}}-(\delta_2-\delta_0)Q^{\text{(nt)}}_0}{\delta_2-\delta_1} \triangleq\xi(Q^{\text{(nt)}}_0),
\end{array}
\end{equation}
where $\delta=\frac{Q^{(t)}}{Q^{\text{(nt)}}}$.
Then, taking the overall QBER $E^{(t)}=\sum^{\infty}_{n=0}\frac{Q^{(t)}_n e^{(t)}_n}{Q^{(t)}}$ with triggering and the one $E^{\text{(nt)}}=\sum^{\infty}_{n=0}\frac{Q^{\text{(nt)}}_n e^{\text{(nt)}}_n}{Q^{\text{(nt)}}}$ without triggering into account, one can derive a upper bound for the single-photon error rate \cite{Pas-Adachi07}:
\begin{equation}
\begin{array}{lll}
e_1 \leq \min(\frac{2\delta E^{(t)}Q^{\text{(nt)}}-\delta_0Q^{\text{(nt)}}_0}{2\delta_1\xi(Q^{\text{(nt)}}_0)}, \frac{2E^{\text{(nt)}}Q^{\text{(nt)}}-Q^{\text{(nt)}}_0}{2\xi(Q^{\text{(nt)}}_0)})\triangleq\epsilon(Q^{\text{(nt)}}_0),
\end{array}
\end{equation}
where $0\leq Q^{\text{(nt)}}_0\leq \min(2E^{(t)}Q^{\text{(nt)}}(\delta/\delta_0), 2E^{\text{(nt)}}Q^{\text{(nt)}})$.

Takeing both of the keys derived from the triggered events and nontriggered events into consideration, and applying the GLLP formula \cite{GLLP04}, one can obtain the final key rate which is shown by Eqs.(13) and (14) in Ref. \cite{Pas-Adachi07}.

\section{Passive decoy-state protocol with finite-length keys}
Due to the effect of finite-size data sets in real-life experiments, there exist various fluctuations in the parameter-estimation step \cite{decoy-Ma}. For a SPDCS, it is proved that the AYKI protocol actually always holds with whatever intensity fluctuation of pump light \cite{Wang10}. Hence, in this paper, we mainly consider the influence of the finite-size effect on the estimation of single-photon yield, single-photon error rate, and phase error rate.

\subsection{Phase error rate}

Here, the phase errors, an argument arising from the Shor--Preskill formalism \cite{Lo99}, means that the maximal virtual errors come from the activity of smart eavesdroppers. It can not be directly measured in experiment and, in the case of a finite-size data set, has to be estimated via a random-sampling theory according to the observed bit errors. In this paper, we apply the interval estimation based on the straightforward bounds \cite{Hayashi12} from an approaching technique for the hypergeometric distribution. It should be noted that this estimation is in accordance with the security criteria based on trace distance and, most importantly, is proved to be tighter than the one in Ref. \cite{Tomamichel12} and more stringent than the one in Ref. \cite{Fung10}.

\textit{Lemma 3 (straightforward bound)}. Let $n$, $l$ and $c$ be the sifted bits, sample bits and observed error bits, respectively. Suppose the final keys of the QKD protocol are $\varepsilon_{\text{sec}}$-secret, then their phase error rate $e_p$ is given by \cite{Hayashi12}
\begin{equation}
\begin{array}{lll}
e_p=\frac{(n+l)\hat{e}(c+2)-le_{\text{ob}}(c+2)}{n}\triangleq g(e_{\text{ob}}(c)),
\end{array}
\end{equation}
with
\begin{equation}
\begin{array}{lll}
\hat{e}(c)=\frac{e_{\text{ob}}(c)+2\tau+2\sqrt{\tau\{e_{\text{ob}}(c)[1-e_{\text{ob}}(c)]+\tau\}}}{1+4\tau},\\
\tau=\frac{\omega^2n}{4l(n+l-1)},\\
e_{\text{ob}}(c)=c/l,
\end{array}
\end{equation}
where $\omega$ is chosen satisfying
\begin{equation}
\begin{array}{lll}
\sqrt{\frac{n+l}{n}}\sqrt{\frac{\omega^2+2\pi}{2}}e^{\nu}\Phi(\omega)\leqslant \frac{1}{16}{\varepsilon_{\text{sec}}}^2.
\end{array}
\end{equation}
Here, $\nu=\frac{1}{6n}+\frac{1}{12}$ and $\Phi(\omega)=\frac{1}{\sqrt{2\pi}}\int_{\omega}^{\infty}{\text{exp}(\frac{-y^2}{2})\text{dy}}$.

\subsection{single-photon yield}

In the case of finite-length keys, the yield of the triggered $n$-photon events are no longer equal to that of the nontriggered ones, i.e.,
\begin{equation}
\begin{array}{lll}
Y^{(t)}_n \neq Y^{\text{(nt)}}_n.
\end{array}
\end{equation}
However, by the theory of probability statistics, there certainly exist relations between the two parts in concrete confidence regions. This means, that the yield of the triggered $n$-photon events is $\xi$ close to that of the nontriggered ones, which corresponds to the two parts being equal except with a probability of $\epsilon_n$. Here, we consider the bound widely used in finite-key QKD and first introduce the following lemma into estimating the relation between the yields \cite{Tomamichel12,Mertz13}:

\textit{Lemma 4}. Let $\epsilon_n>0$ and $n_1, n_2>0$. Let $\rho^{n_1}$ and  $\rho^{n_2}$ be the quantum state of the triggered and nontriggered $n$-photon events, respectively. They are both permutation-invariant quantum states, and let $\mathcal{E}$ be a positive-operator-value measure (POVM) on $\mathcal{H}_{AB}$ which outputs the yield and quantum bit error rate, where $\rho^{n_1+n_2}\in\mathcal{S}(\mathcal{H}_{AB}^{\otimes n_1+n_2})$. Let $\textbf{Y}^{(t)}_n$ and $\textbf{Y}^{\text{(nt)}}_n$ be the frequency distribution of the measurement events, e.g., the yield, when applying the measurement $\mathcal{E}^{n_1}$ and $\mathcal{E}^{n_2}$, respectively. Then, for any element $Y^{(t)}_n$ and $Y^{(\text{nt})}_n$ from $\textbf{Y}^{(t)}_n$ and $\textbf{Y}^{\text{(nt)}}_n$ except with probability $\epsilon_n$,
\begin{equation}
\label{Sta-1}
\begin{array}{lll}
\frac{1}{2}\parallel Y^{(t)}_n-Y^{\text{(nt)}}_n\parallel \leqslant \xi(\epsilon_n, n_1, n_2),
\end{array}
\end{equation}
with $\xi(\epsilon_n, n_1, n_2)=\sqrt{\frac{(n_1+n_2)(n_1+1)\ln(1/\epsilon_n)}{8{n_1}^2 n_2}}$, where $n_1$ and $n_2$ are the number of $n$-photon triggered events and $n$-photon nontriggered events, respectively, chosen for parameter estimation.

Note that the overall detection rate with triggering and without triggering are expressed respectively by
\begin{equation}
\label{Q-T}
\begin{array}{lll}
Q^{(t)}=\sum\limits^{\infty}_{n=0}Q^{(t)}_n=\sum\limits^{\infty}_{n=0}Y^{(t)}_n p_n \gamma_n,
\end{array}
\end{equation}

\begin{equation}
\label{Q-NT}
\begin{array}{lll}
Q^{\text{(nt)}}=\sum\limits^{\infty}_{n=0}Q^{(\text{nt})}_n=\sum\limits^{\infty}_{n=0}Y^{\text{(nt)}}_n p_n (1-\gamma_n).
\end{array}
\end{equation}
Equation (\ref{Q-NT}) is multiplied by $\delta_2$ and we obtain
\begin{equation}
\label{F-1}
\begin{array}{lll}
\delta_2Q^{\text{(nt)}}=\delta_2Q^{(\text{nt})}_0+\delta_2Q^{(\text{nt})}_1+\delta_2\sum\limits^{\infty}_{k=2}Q^{(\text{nt})}_k .
\end{array}
\end{equation}
From Eq.(\ref{Sta-1}), we can find that $Y^{\text{(nt)}}_n\leqslant Y^{(t)}_n+2\xi_n$, where $\xi_n=\xi(\epsilon_n, n_1, n_2)$. Then, the third term of the right-hand side of the above equation satisfies
\begin{equation}
\label{F-2}
\begin{array}{lll}
\delta_2\sum\limits^{\infty}_{k=2}Q^{(\text{nt})}_k\leqslant \delta_2\sum\limits^{\infty}_{k=2}(Y^{(t)}_k+2\xi_k)p_k (1-\gamma_k)\\
\quad\quad\quad\quad\quad \leqslant \sum\limits^{\infty}_{k=2}(Y^{(t)}_k+2\xi_k)p_k\gamma_k \\
\quad\quad\quad\quad\quad =\sum\limits^{\infty}_{k=2}Q^{(t)}_k+2\sum\limits^{\infty}_{k=2}\xi_k p_k\gamma_k,
\end{array}
\end{equation}
where $\sum\limits^{\infty}_{k=2}Q^{(t)}_k=Q^{(t)}-Q^{(t)}_0-Q^{(t)}_1$. Hence, from Eq.(\ref{Q-NT}), one can obtain
\begin{equation}
\label{F-3}
\begin{array}{lll}
\delta_2(Q^{\text{(nt)}}-Q^{(\text{nt})}_0-Q^{(\text{nt})}_1)\\
 \leqslant Q^{(t)}-Q^{(t)}_0-Q^{(t)}_1+2\sum\limits^{\infty}_{k=2}\xi_k p_k\gamma_k \\
 \leqslant Q^{(t)}+(2\xi_1-Y^{\text{(nt)}}_1)p_1\gamma_1+(2\xi_0-Y^{\text{(nt)}}_0)p_0\gamma_0\\
 \quad+2\sum\limits^{\infty}_{k=2}\xi_k p_k\gamma_k\\
 = Q^{(t)}-\delta_0Q^{(\text{nt})}_0-\delta_1Q^{(\text{nt})}_1+2\sum\limits^{\infty}_{k=0}\xi_k p_k\gamma_k.
\end{array}
\end{equation}

We thus obtain a minimum value of $Q^{(\text{nt})}_1$ as a function of $Q^{(\text{nt})}_0$:
\begin{equation}
\label{F-4}
\begin{array}{lll}
Q^{(\text{nt})}_1\geqslant \frac{\delta_2 Q^{\text{(nt)}}-Q^{(t)}-(\delta_2-\delta_0)Q^{(\text{nt})}_0-2\sum\limits^{\infty}_{k=0}\xi_k p_k\gamma_k}{\delta_2-\delta_1}.
\end{array}
\end{equation}
Let $p_{\text{pe}}$ be the probability of choosing a pulse from the SPDC process as the sample bits used for parameter estimation. Then, if we assume $\epsilon_{\text{\text{pe}}}=\epsilon_0=\epsilon_1=\epsilon_2=\cdots$ and note that $k_1=Np_{\text{pe}}p_k\gamma_k$ and $k_2=Np_{\text{pe}}p_k(1-\gamma_k)$ in $\xi_k=\xi(\epsilon_k, k_1, k_2)$, the above bound can be further represented by
\begin{equation}
\label{F-5}
\begin{array}{lll}
\frac{Q^{(\text{nt})}_1}{Q^{\text{(nt)}}}\geqslant \frac{[(\delta_2-\delta)-(\delta_2-\delta_0)x-\chi]}{\delta_2-\delta_1}\triangleq \zeta(x),
\end{array}
\end{equation}
where $N$ denotes the number of total pulses emitted from the SPDC process, $x=\frac{Q^{\text{(nt)}}_0}{Q^{\text{(nt)}}}$ and $\chi=\frac{1}{Q^{\text{(nt)}}}\sqrt{\frac{\ln{(1/\epsilon_{\text{\text{pe}}})}}{2Np_{\text{pe}}}}\sum\limits^{\infty}_{k=0}\sqrt{\delta_k p_k}$. From Eq.(\ref{Sta-1}), one can also find that $Y^{(t)}_1 \geqslant Y^{\text{(nt)}}_1-2\xi_1$. Therefore, we can also obtain a lower bound for $Q^{(t)}_1$:
\begin{equation}
\label{F-6}
\begin{array}{lll}
Q^{(t)}_1\geqslant \delta_1Q^{\text{(nt)}}_1-\chi_1\geqslant \delta_1Q^{\text{(nt)}}\zeta(x)-\chi_1,
\end{array}
\end{equation}
where $\chi_1=\sqrt{\frac{\delta_1p_1\ln{(1/\epsilon_{\text{\text{pe}}})}}{2Np_{\text{pe}}}}$.

\subsection{single-photon error rate}

The overall quantum bit error rate for the triggered events and nontriggered events can be represented, respectively, by
\begin{equation}
\label{E-1}
\begin{array}{lll}
Q^{(t)}E^{(t)}=\sum\limits^{\infty}_{n=0}Q^{(t)}_n e^{(t)}_n=\sum\limits^{\infty}_{n=0}Y^{(t)}_n e^{(t)}_n p_n \gamma_n,
\end{array}
\end{equation}

\begin{equation}
\label{E-2}
\begin{array}{lll}
Q^{\text{(nt)}}E^{\text{(nt)}}=\sum\limits^{\infty}_{n=0}Q^{(\text{nt})}_n e^{\text{(nt)}}_n=\sum\limits^{\infty}_{n=0}Y^{\text{(nt)}}_n e^{\text{(nt)}}_n p_n (1-\gamma_n).
\end{array}
\end{equation}

From Eqs.(\ref{Sta-1}), (\ref{F-6}) and (\ref{E-1}) with $e^{(t)}_0=\frac{1}{2}$, an upper bound on $e^{(t)}_1$ is given by
\begin{equation}
\label{E-3}
\begin{array}{lll}
e^{(t)}_1\leqslant \frac{Q^{(t)}E^{(t)}-Q^{(t)}_0e^{(t)}_0}{Q^{(t)}_1}\leqslant \frac{2Q^{(t)}E^{(t)}-Q^{(t)}_0}{2(\delta_1Q^{\text{(nt)}}_1-\chi_1)}\\
\qquad\qquad\qquad\qquad \leqslant \frac{2Q^{(t)}E^{(t)}-\delta_0Q^{(\text{nt})}_0+\chi_0}{2[\delta_1Q^{\text{(nt)}}\zeta(x)-\chi_1]}\\
\qquad\qquad\qquad\qquad \leqslant
\frac{2\delta E^{(t)}-\delta_0 x+\chi_0/ Q^{\text{(nt)}}}{2\delta_1\zeta(x)-2\chi_1/ Q^{\text{(nt)}}}\triangleq \mathcal{W}_t(x),
\end{array}
\end{equation}
where $\chi_0=\sqrt{\frac{\delta_0p_0\ln{(1/\epsilon_{\text{\text{pe}}})}}{2Np_{\text{pe}}}}$.

Similarly, from Eqs.(\ref{Sta-1}), (\ref{F-5}) and (\ref{E-2}) with $e^{\text{(nt)}}_0=\frac{1}{2}$, one can have an upper bound on $e^{\text{(nt)}}_1$ , which is shown by
\begin{equation}
\label{E-4}
\begin{array}{lll}
e^{\text{(nt)}}_1\leqslant \frac{Q^{\text{(nt)}}E^{\text{(nt)}}-Q^{\text{(nt)}}_0e^{\text{(nt)}}_0}{Q^{\text{(nt)}}_1}\leqslant \frac{2E^{\text{(nt)}}-x}{2\zeta(x)} \triangleq \mathcal{W}_{\text{nt}}(x)
\end{array}
\end{equation}

\subsection{Secret key length}

If we consider the secret key only from the triggered events and apply Lemma 2, a $\varepsilon_{\text{sec}}$-secret key of length $\ell$ can be given by
\begin{equation}
\label{S-1}
\begin{array}{lll}
\lfloor H^{\nu}_{\min}(X_A^{(t)}|E')-2\log_2{\frac{1}{2\overline{\nu}}} \rfloor,
\end{array}
\end{equation}
where $X_A^{(t)}$ is the raw key extracted from the triggered events, $\nu+\overline{\nu}\leq \varepsilon_{\text{sec}}$ with $\nu$ and $\overline{\nu}$ chosen to be  proportional to $\frac{\varepsilon_{\text{sec}}}{p_{\text{pass}}}$.

Then, applying the results of Ref. \cite{Lim13}, the length of secret key from the triggered events can be represented by
\begin{equation}
\label{S-2}
\begin{array}{lll}
\lfloor n_0^{(t)}+n_1^{(t)}(1-h(e_p^{(t)}))-\lambda_{EC}^{(t)}-6\log_2\frac{10}{\varepsilon_{\text{sec}}}-\log_2\frac{2}{\varepsilon_{\text{cor}}} \rfloor,
\end{array}
\end{equation}
where $h(x):=-x\log_2x-(1-x)\log_2(1-x)$ is the binary entropy function, $n_0^{(t)}\geqslant NQ^{\text{(nt)}}(\delta_0x-\frac{\chi_0}{Q^{\text{(nt)}}})$ with $\chi_0=\sqrt{\frac{\delta_0p_0\ln{(1/\epsilon_{\text{\text{pe}}})}}{2N}}$, $n_1^{(t)}\geqslant NQ^{\text{(nt)}}[\delta_1\zeta(x)-\frac{\chi_1}{Q^{\text{(nt)}}}]$, $e_p^{(t)}$ denotes the phase error rate which is calculated by Lemma 3, $\lambda_{EC}^{(t)}=NQ^{(t)}f_{EC}h(E^{(t)})$. It should be noted that $\varepsilon_{\text{sec}}=9\varepsilon+\epsilon_{\text{\text{pe}}}$ where $\epsilon_{\text{\text{pe}}}$ is the failure probability of estimating the single-photon yield and error rate mentioned in the previous subsection. Let $\epsilon_{\text{\text{pe}}}=\varepsilon$, then $\varepsilon_{\text{sec}}=10\varepsilon$, which is different from Ref. \cite{Lim13}. $\varepsilon_{\text{cor}}$ is the security parameter of the error-verification step. Hence, the length of the secret key from the triggered events can be shown as
\begin{equation}
\label{S-22}
\begin{array}{lll}
\ell_{T}= \min\limits_x \{N(1-p_{\text{pe}})Q^{\text{(nt)}}[\delta_0x-\frac{\chi_0}{Q^{\text{(nt)}}}\\
\quad\quad+(\delta_1\zeta(x)-\frac{\chi_1}{Q^{\text{(nt)}}})(1-h(e_p^{(t)}))]\}\\
\quad\quad -N(1-p_{\text{pe}})Q^{(t)}f_{EC}h(E^{(t)})\\
\quad\quad -6\log_2\frac{10}{\varepsilon_{\text{sec}}}-\log_2\frac{2}{\varepsilon_{\text{cor}}},
\end{array}
\end{equation}
where $\chi_i=\sqrt{\frac{\delta_ip_i\ln{(1/\epsilon_{\text{pe}})}}{2Np_{\text{pe}}}}$ with $i=0$ or $i=1$, $p_{\text{pe}}$ is the probability of choosing a pulse from the SPDC process as the sample events used for parameter estimation and $e_p^{(t)}=g(\mathcal{W}_t(x))$. The minimum is numerically taken over the range $0\leqslant x\leqslant \min\{2E^{(t)}\delta/\delta_0, 2E^{\text{(nt)}}\}$.

However, if we also take the secret key from the nontriggered events into account when the error reconciliation is separately
applied to the triggered events and to the nontriggered events, but the privacy amplification is applied together, Eq.(\ref{S-1}) no longer holds true and we shall recalculate the length of the secret key by
\begin{equation}
\label{S-3}
\begin{array}{lll}
\lfloor H^{\nu}_{\min}(X_A^{(t)}X_A^{\text{(nt)}}|E^{(t)'}E^{\text{(nt)}'})-2\log_2{\frac{1}{2\overline{\nu}}} \rfloor,
\end{array}
\end{equation}
where $X_A^{(t)}$ and $X_A^{\text{(nt)}}$ are the raw key extracted from the triggered and nontriggered events respectively, $E^{(t)'}$ and $E^{\text{(nt)}'}$ are the information that Eve gathers on $X_A^{(t)}$ and $X_A^{\text{(nt)}}$, respectively, up to the error verification step. In the following, we will show how to estimate a lower bound of the left term in Eq.(\ref{S-3}).

By Lemma 1, we have that
\begin{equation}
\label{S-4}
\begin{array}{lll}
H^{\nu}_{\min}(X_A^{(t)}X_A^{\text{(nt)}}|E^{(t)'}E^{\text{(nt)}'})\\
\quad\quad\quad\geqslant H^{\nu_2}_{\min}(X_A^{(t)}|X_A^{\text{(nt)}}E^{(t)'}E^{\text{(nt)}'})\\
\quad\quad\quad\quad +H^{\nu_3}_{\min}(X_A^{\text{(nt)}}|E^{(t)'}E^{\text{(nt)}'})-f(\nu_1)\\
\quad\quad\quad =H^{\nu_2}_{\min}(X_A^{(t)}|E^{(t)'})+H^{\nu_3}_{\min}(X_A^{\text{(nt)}}|E^{\text{(nt)}'})-f(\nu_1),
\end{array}
\end{equation}
where
\begin{equation}
\label{S-5}
\begin{array}{lll}
f(\nu_1)=\log_2(2/\nu_1^2),\\
\nu=\nu_1+2\nu_2+\nu_3, \\ H^{\nu_2}_{\min}(X_A^{(t)}|E^{(t)'})\geqslant H^{\nu_2}_{\min}(X_A^{(t)}|E^{(t)})-\lambda_{EC}^{(t)}-\log_2(2/\varepsilon_{\text{cor}}), \\ H^{\nu_3}_{\min}(X_A^{\text{(nt)}}|E^{\text{(nt)}'})\\
\quad\quad\quad \geqslant H^{\nu_3}_{\min}(X_A^{\text{(nt)}}|E^{\text{(nt)}})-\lambda_{EC}^{\text{(nt)}}-\log_2(2/\varepsilon_{\text{cor}}).
\end{array}
\end{equation}
In the above equations, $E^{(t)}$ and $E^{\text{(nt)}}$ denote the remaining quantum information that Eve has on $X_A^{(t)}$ and $X_A^{\text{(nt)}}$, respectively, after the error correction and error verification steps. According to the analysis of Ref. \cite{Lim13}, the terms $H^{\nu_2}_{\min}(X_A^{(t)}|E^{(t)})$ and $H^{\nu_3}_{\min}(X_A^{\text{(nt)}}|E^{\text{(nt)}})$ in Eq.(\ref{S-5}) can be lower bounded by the generalized chain-rule result (Lemma 1 \cite{Vitanov13}) and the uncertainty relation for smooth entropies \cite{Tomamichel12}. Precisely, they are given by
\begin{equation}
\label{S-6}
\begin{array}{lll}
H^{\nu_2}_{\min}(X_A^{(t)}|E^{(t)})\\
\quad\quad\quad \geqslant n_0^{(t)}+n_1^{(t)}(1-h(e_p^{(t)}))-\log_2\frac{2}{(\alpha_2\alpha_3)^2}, \\ H^{\nu_3}_{\min}(X_A^{\text{(nt)}}|E^{\text{(nt)}})\\
\quad\quad\quad \geqslant n_0^{\text{(nt)}}+n_1^{\text{(nt)}}(1-h(e_p^{\text{(nt)}}))-\log_2\frac{2}{(\alpha_5\alpha_6)^2},
\end{array}
\end{equation}
where $\nu_2=2\alpha_1+\alpha_2+\alpha_3$ and $\nu_3=2\alpha_4+\alpha_5+\alpha_6$.

Combing Eqs.(\ref{S-3}-\ref{S-6}), the final secret key from both the triggered and nontriggered events is said to be $\varepsilon_{\text{sec}}$-secret if its length $\ell_{B}$ is chosen by
\begin{equation}
\label{S-7}
\begin{array}{lll}
\ell_{B} \geqslant
n_0^{(t)}+n_0^{\text{(nt)}}+n_1^{(t)}(1-h(e_p^{(t)}))+n_1^{\text{(nt)}}(1-h(e_p^{\text{(nt)}}))\\
\quad\quad-\lambda_{EC}^{(t)}-\lambda_{EC}^{\text{(nt)}}-\log_2{\frac{2}{\nu_1^2}}-\log_2\frac{4}{\alpha_{p}^2},
\end{array}
\end{equation}
with
\begin{equation}
\label{S-8}
\begin{array}{lll}
\varepsilon_{\text{sec}}=\nu+\overline{\nu}+\epsilon_{\text{pe}}\\
\quad\quad=\nu_1+2(2\alpha_1+\alpha_2+\alpha_3)+2\alpha_4+\alpha_5+\alpha_6+\overline{\nu}+\epsilon_{\text{pe}},\\
\alpha_{p}=\alpha_2\alpha_3\alpha_5\alpha_6\varepsilon_{\text{cor}}\overline{\nu},
\end{array}
\end{equation}
where $\epsilon_{\text{pe}}$ is the failure probability of estimating the single-photon yield and error rate.

For evaluation, we set each error term in Eq.(\ref{S-10}) to a common value $\varepsilon$ and let $\epsilon_{\text{pe}}=\varepsilon$. Therefore, the secrecy for the key obtained from both the triggered events and nontriggered events is $\varepsilon_{\text{sec}}=15\varepsilon$. Then, considering the bounds of single-photon yield and error rate given in the previous subsections, $\ell_B$ can be obtained as the following
\begin{equation}
\label{S-9}
\begin{array}{lll}
\ell_{B}=
\min\limits_x \{N(1-p_{\text{pe}})Q^{\text{(nt)}}[(\delta_0x+x-\frac{\chi_0}{Q^{\text{(nt)}}})\\
\quad\quad+(\delta_1\zeta(x)-\frac{\chi_1}{Q^{\text{(nt)}}})(1-h(e_p^{(t)}))+\zeta(x)(1-h(e_p^{\text{(nt)}}))]\}\\
\quad\quad-N(1-p_{\text{pe}})Q^{(t)}f_{\text{EC}}h(E^{(t)})\\
\quad\quad-N(1-p_{\text{pe}})Q^{\text{(nt)}}f_{\text{EC}}h(E^{\text{(nt)}})\\
\quad\quad-2\log_2\frac{15}{\varepsilon_{\text{sec}}}-1-10\log_2\frac{15}{\varepsilon_{\text{sec}}}-\log_2\frac{4}{\varepsilon_{\text{cor}}},
\end{array}
\end{equation}
where
\begin{equation}
\label{S-10}
\begin{array}{lll}
\chi_i=\sqrt{\frac{\delta_ip_i\ln{(15/\varepsilon_{\text{sec}})}}{2Np_{\text{pe}}}} \quad\text{with} \quad i=0 \quad\text{or} \quad 1,\\
e_p^{(t)}=g(\mathcal{W}_t(x)),\\
e_p^{\text{(nt)}}=g(\mathcal{W}_{\text{nt}}(x)).
\end{array}
\end{equation}

In Eqs.(\ref{S-9}) and (\ref{S-10}), the minimum is numerically taken over the range $0\leqslant x\leqslant \min\{2E^{(t)}\delta/\delta_0, 2E^{\text{(nt)}}\}$ \cite{Pas-Adachi07}.

To conclude, the length of the final secret key can be given as $\ell=\max\{\ell_{T},\ell_{B}\}$.
\section{numerical simulation}
In this section, by assuming a fiber-based channel model, we numerically show the performance of our protocol with finite-length key. Let $\eta_c=10^{-\alpha L/10}$ being the fiber transmission with $\alpha=0.2$ dB/km the attenuation coefficient, $\eta_B$ the quantum efficiency of Bob's detectors and $\eta\equiv\eta_c\eta_B$. For better comparison, we borrow experimental parameters from Ref. \cite{Lim13}, which assumes that Bob uses an active measurement setup with two single-photon detectors with total detection efficiency $\eta_B=0.1$ and dark-count probability $p_d=6\times10^{-7}$. On the sender's side, we assume Alice uses a SPDCS and a typical silicon avalanche photodiode as threshold detector with $d_A=10^{-6}$ and $\eta_A=0.5$. The numerical parameters used are listed in Table I.
\begin{table}[!t]
\label{Table I}
\caption{List of experimental parameters for simulations: $\alpha$ is the loss coefficient of the fiber, $f_{\text{EC}}$ is the error-correction efficiency, $\eta_B$ is the detection efficiency of Bob's detectors, $e_d$ is the error rate due to optical errors, which is the probability that a photon sent from Alice hits the erroneous detector, $p_d$ is the background dark-count rate of Bob's detectors, $\eta_A$ and $d_A$ are the detection efficiency and dark count rate of Alice's detector, respectively.}
\tabcolsep=6pt
\begin{tabular*}{86mm}{lcccccc}\hline\hline
$ \alpha$(dB/km) & $f_{\text{EC}}$  & $\eta_B$ & $e_d$ &        $p_d$       &   $d_A$   & $\eta_A$ \\ \hline
      0.20       &      1.16        &    0.1   & 0.005 & $6\times10^{(-7)}$ & $10^{-6}$ &   0.5    \\ \hline\hline
\end{tabular*}
\end{table}

For the average overall gain $Q^{(t)}$ and $Q^{\text{(nt)}}$, also the average quantum bit error rate (QBER) $E^{(t)}$ and $E^{\text{(nt)}}$, they can be directly measured in the experiment. In this paper, for simulation purpose, we neglect the finite size effect in the calculation of the average overall gain and QBER. Then, according to the channel model, it is given that
\begin{equation}
\label{M-1}
\begin{array}{lll}
Q^{(t)}=\sum\limits_{n=0}^{\infty}p_n\gamma_n[1-(1-\eta)^n(1-p_d)^2], \\ Q^{\text{(nt)}}=\sum\limits_{n=0}^{\infty}p_n(1-\gamma_n)[1-(1-\eta)^n(1-p_d)^2], \\
E^{(t)}=\frac{1}{2Q^{(t)}}\sum\limits_{n=0}^{\infty}p_n\gamma_n\{1-(1-\eta)^n(1-p_d)^2 \\
\quad\quad\quad-(1-p_d)[(1-\eta e_d)^n-(1-\eta+\eta e_d)^n]\}, \\
E^{\text{(nt)}}=\frac{1}{2Q^{(nt)}}\sum\limits_{n=0}^{\infty}p_n(1-\gamma_n)\{1-(1-\eta)^n(1-p_d)^2 \\
\quad\quad\quad-(1-p_d)[(1-\eta e_d)^n-(1-\eta+\eta e_d)^n]\},
\end{array}
\end{equation}
with $p_n=\frac{\mu^n}{(1+\mu)^{n+1}}$ and $\gamma_n=1-(1-d_A)(1-\eta_A)^n$. The summations in Eq.(\ref{M-1}) can be solved mathematically. However, for simplicity, we do not give their expressions here.

\begin{figure}[!t]\centering
\resizebox{9.3cm}{7cm}{
\includegraphics{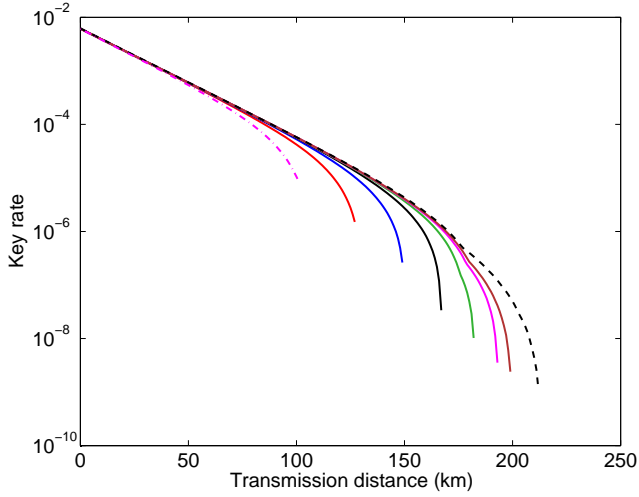}}
\caption{(Color online) Secret key rate vs transmission distance. The secret key rates from left to right are numerically optimized for fixed number of total pulses from the SPDC process $N=10^j$ with $j=9,10,\ldots,15$. The dashed curve denotes to the asymptotic secret key rate calculated from Eqs.(9)-(14) in Ref. \cite{Pas-Adachi07}, i.e., the key rate of the AYKI protocol with keys of infinite length; The intensity $\mu$ of the SPDCS and the probability $p_{\text{pe}}$ of sample events in the total pulses are numerically chosen to be optimal for different transmission distances.}
\end{figure}

In our simulations, the key's secrecy $\varepsilon_{\text{sec}}$ and correctness $\varepsilon_{\text{cor}}$ are set to be $10^{-10}$ and $10^{-12}$, respectively. For the estimation of the phase error rate, we assume $n=l\geqslant 125$. Note that the analysis in Ref. \cite{Hayashi12} is based on the QKD protocol with an ideal single-photon source. However, in our paper, a practical SPDCS is used in our protocol, which is within reach of current technology. Hence, the sifted key bits $n$ should be replaced by the fraction bits of the single-photon contribution, that is, $n=N(1-p_{\text{pe}})Q_1$. Here, $Q_1$ represents the gain from the single-photon detections. Then, for the estimate of phase error rate in Eqs.(\ref{S-22}) and (\ref{S-9}), i.e., $e_p^{(t)}$ and $e_p^{\text{(nt)}}$, we shall set their $n$ be $N(1-p_{\text{pe}})Q^{(t)}_1$ and $N(1-p_{\text{pe}})Q^{\text{(nt)}}_1$, respectively. Likewise, the number of sample bits used for parameter estimation, i.e., $l^{(t)}$ and $l^{\text{(nt)}}$ , are set to be $Np_{\text{pe}}Q^{(t)}_1$ and $Np_{\text{pe}}Q^{\text{(nt)}}_1$, respectively. Under these conditions, we apply an optimization about the secret key rate $R=\ell/(2N)$ over $\{x, \mu, p_{\text{pe}}\}$ given that the set $\{\varepsilon_{\text{sec}}, \varepsilon_{\text{cor}}, e_d, \eta_B, p_d, d_A, \eta_A, f_{\text{EC}}, Q^{(t)}, Q^{\text{(nt)}}, E^{(t)}, E^{\text{(nt)}}\}$ is fixed.

\begin{figure}[!t]\centering
\resizebox{9.3cm}{7cm}{
\includegraphics{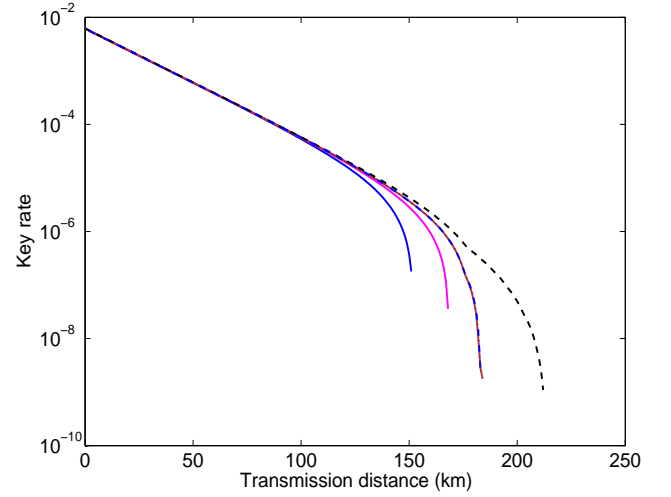}}
\caption{(Color online) Secret key rate vs transmission distance. $N$ is fixed to be $10^{13}$. For different $p_{\text{pe}}=0.01, 0.1, 0.9$, the solid lines (from left to right) are still plotted with intensity $\mu$ chosen optimally for different transmission distances. The dash-dotted curve denotes to the secret key rate when $p_{\text{pe}}=0.96$ and the dashed curve still corresponds to the asymptotic secret key rate.}
\end{figure}

In Fig.1, the numerically optimized secret key rates from left to right are obtained by Eqs. (\ref{S-9}) and (\ref{S-10}) for a fixed number of total pulses $N=10^j$ with $j=9,10,\ldots,15$, respectively. It should be noted that, so as to reach the distance of 100 km, $N$ should be at least $10^9$. When we consider the postprocessing block size $n_X$ mentioned in Ref. \cite{Lim13}, the least requirement corresponds to $n_X\geqslant N(1-p_{\text{pe}})R\geqslant 10^6$, larger than the one with a WCS by Lim et al. \cite{Lim13}. But within the distance of 50 km for $N=10^9$, the secret key rate calculated by our method can be higher than $5.397\times 10^{-4}$, which is better than the one using a WCS (lower than $10^{-4}$ by Fig.1 of \cite{Lim13}). And when one fixes $N$ to $10^{13}$ corresponding to a postprocessing block size of $10^{9}$, the maximal transmission distance can reach to 167 km, which is getting close to 180 km with the WCS under the same conditions. Most importantly, from Fig.1, one can find that the maximal transmission distance of our method for $N=10^{13}$, $10^{14}$ and $10^{15}$ can be longer than 182 km, which performs better than the case using WCS for a postprocessing block size of $10^9$.

In Fig.2, we also simulate the secret key rates for different probabilities of $p_{\text{pe}}$. Without loss of generality, when we fix $N$ to be $10^{13}$, one can see that the the smaller value of $p_{\text{pe}}$ results a lower secret key generation rate. When $p_{\text{pe}}$ is larger than $0.9$ (e.g. $p_{\text{pe}}=0.96$ ), the maximal transmission distance is almost unchanged, representing the optimality of $p_{\text{pe}}$ at $0.9$. And our simulation is likely to help experimentalists to improve the performance of their QKD experiments using SPDCS.

\section{conclusion}
In conclusion, we put forward a passive decoy-state protocol in the finite-size effect. For $\varepsilon-\text{secure}$ secret keys with finite length, the bound for estimating the single-photon contributions, single-photon errors and the length of final keys are presented. From numerical simulations, we remark that the QKD using a SPDCS performs as well as that using WCS. Furthermore, we conclude that our  passive decoy-state protocol with a finite-length key can reach a higher secret transmission distance than that using WCS. And our protocol can certainly be considered as a choice for the practical experiment of QKD using SPDCS.

We notice that a recent study by Krapick et al. \cite{Krapick14} proposed a kind of SPDCS with bright intensity, which can be applied in our passive decoy-state protocol. It will be attractive and interesting to analyze the performance of QKD using this source. We shall concentrate on this issue in future works.
\section*{ACKNOWLEDGMENTS}
 The authors gratefully acknowledge the financial support from the National Basic Research Program of China (Grant No. 2013CB338002) and the National Natural Science Foundation of China (Grants No.11304397, No.61101137, No.61201239 and No. 61205118).



\begin{thebibliography}{10}

\bibitem{BB84} C. H. Bennett, G. Brassard, in Proceedings IEEE Int. Conf. on Computers, Systems and Signal Processing, Bangalore, India (IEEE, New York, 1984), pp. 175-179, (1984)

\bibitem{Lo99} H.-K. Lo, H. F. Chau, Science 283, 5410 (1999); P. W. Shor and J. Preskill Phys. Rev. Lett. 85, 441 (2000).
\bibitem{GLLP04} D. Gottesman, H.-K. Lo, Norbert L\"{u}kenhaus, and John Preskill, Quant. Inf. Comp. 4, 325 (2004)
\bibitem{Tomamichel11} M. Tomamichel and R. Renner, Phys. Rev. Lett. 106, 110506 (2011).
\bibitem{Guo12} S. Wang, W. Chen, J. F. Guo, Z. Q. Yin, et al., Opt. Lett. 37, 1008 (2012).
\bibitem{Zeilinger12} X. S. Ma et al., Nature 489, 269 (2012).
\bibitem{Bacco13} D. Bacco, M. Canale, N. Laurenti, G. Vallone, and P. Villoresi, Nat. Commun. 4 (2013).
\bibitem{Pan13} J.-Y. Wang et al., Nat. Photonics 7, 387 (2013).
\bibitem{Shield13} B. Frohlich, J. F. Dynes, M. Lucamarini, A. W. Sharpe, Z. Yuan, and A. J. Shields, Nature 501, 69 (2013).

\bibitem{Scarani09} V. Scarani, H. Bechmann-Pasquinucci, N. J. Cerf, M. Dusek, N. L\"{u}tkenhaus, and M. Peev, Reviews of Modern Physics 81, 1301 (2009).

\bibitem{Hacking1} Y. Zhao, C.-H. F. Fung, B. Qi, C. Chen and H.-K. Lo. Phys. Rev. A 78,042333(2008).
\bibitem{Hacking2} L. Lydersen, C. Wiechers, C. Wittmann, D. Elser, J. Skaar, and V. Makarov, Nat. Photonics 4, 686 (2010).
\bibitem{Hacking3} H. W. Li et al., Phys. Rev. A 84, 062308 (2011).

\bibitem{DI1} A. Ac\'{\i}n, N. Brunner, N. Gisin, S. Massar, S. Pironio and V. Scarani, Phys. Rev. Lett. 98, 230501 (2007).
\bibitem{DI2} N. Gisin, S. Pironio, and N. Sangouard, Phys. Rev. Lett. 105, 070501 (2010);
\bibitem{DI3} M. Paw{\l}owski and N. Brunner, Phys. Rev. A 84, 010302 (2011).
\bibitem{MDI1} H.-K. Lo, M. Curty, and B. Qi, Phys. Rev. Lett. 108, 130503 (2012).
\bibitem{MDI2} S. L. Braunstein and S. Pirandola, Phys. Rev. Lett. 108, 130502 (2012).
\bibitem{MDI-Lo} K. Tamaki, H.-K. Lo, C.-H. F. Fung, and B. Qi, Phys. Rev. A 85, 042307 (2012).
\bibitem{MDI-Ma} X. F. Ma and M. Razavi, Phy. Rev. A 86, 062319 (2012).
\bibitem{Bao13-1} C. Zhou, W. S. Bao, W. Chen, H. W. Li, Z. Q. Yin, Y. Wang, and Z. F. Han, Phys. Rev. A 88, 052333 (2013).


\bibitem{Hayashi07} M. Hayashi, Phys. Rev. A. 76, 012329 (2007); J.Hasegawa, M. Hayashi, T. Hiroshima and A. Tomita, preprint arXiv:0707.3541 (2007).
\bibitem{Cai09} R. Cai and V. Scarani, New J. Phys. 11, 045024 (2009).
\bibitem{Li09} H.-W. Li, Y.-B. Zhao, Z.-Q. Yin, S.Wang, Z.-F. Han, W.-S. Bao, and G.-C. Guo, Opt. Commun. 282, 4162 (2009).
\bibitem{Song11} T. T. Song, J. Zhang, S. J. Qin and Q. Y.Wen, Quantum Inf. Comput. 11, 374-389 (2011).
\bibitem{Somma13} R. D. Somma and R. J. Hughes, Phys. Rev. A. 87, 062330 (2013).

\bibitem{Renner05} R. Renner, Ph.D. thesis, Swiss Federal Institute of Technology (ETH) Zurich, 2005, \textit{quant-ph}/0512258; R. Renner, Int. J. Quantum Inf. 6, 1 (2008).

\bibitem{Scarani08-1} V. Scarani and R. Renner, in Theory of Quantum Computation, Communication, and Cryptography, edited by Y. Kawano and M. Mosca (Springer, Berlin/Heidelberg, 2008), Vol. 5106 of Lecture Notes in Computer Science, pp. 83¨C95.
\bibitem{Scarani08-2} V. Scarani and R. Renner, Phys. Rev. Lett. 100, 200501 (2008).
\bibitem{Christandl09} M. Christandl, R. K\"{o}nig, and R. Renner, Phys. Rev. Lett. 102, 020504 (2009).
\bibitem{Bratzik11} S. Bratzik, M. Mertz, H. Kampermann, and D. Bru{\ss}, Phys. Rev. A 83, 022330 (2011).
\bibitem{Ng12} N.H.Y. Ng, M. Berta, and S. Wehner, Phys. Rev. A 86, 042315 (2012).
\bibitem{Hayashi12} M. Hayashi and T. Tsurumaru, New. J. Phys. 14 (2012).
\bibitem{Hayashi13} M. Hayashi and R. Nakayama, arXiv:1302.4139v3 (2013).

\bibitem{Tomamichel12} M. Tomamichel, C. C. W. Lim, N. Gisin, and R. Renner, Nat. Commun. 3, 634 (2012).
\bibitem{Mertz13} M. Mertz, H. Kampermann, S. Bratzik, and D. Bru{\ss}, Phys. Rev. A 87, 012315 (2013).
\bibitem{Lim13} C.C.W. Lim, M. Curty, N. Walenta, F. Xu, and H. Zbinden, Phys. Rev. A 89, 022307 (2014).
\bibitem{Curty13} M. Curty, F.H. Xu, W. Cui, C. C. W. Lim, K. Tamaki, and H.K. Lo, Nat. Commun. 5, 3732 (2014).
\bibitem{Bao13-2} Y. Wang, W.S. Bao, H.W. Li, C. Zhou, and Y. Li, Phys. Rev. A 88, 052322 (2013).
\bibitem{Mafu13} M. Mafu, K. Garapo, and F. Petruccione, Phys. Rev. A 88, 062306 (2013).
\bibitem{Vitanov13} A. Vitanov, F. Dupuis, M. Tomamichel and R. Renner, IEEE Trans. Inf. Theory 59, 2603-2612 (2013).

\bibitem{Brassard00} G. Brassard, N. L\"{u}tkenhaus, T. Mor, and B. C. Sanders, Phys. Rev. Lett. 85, 1330 (2000).
\bibitem{decoy03} W.-Y. Hwang, Phys. Rev. Lett. 91, 057901 (2003).
\bibitem{decoy-Lo} H.-K. Lo, X. Ma, and K. Chen, Phys. Rev. Lett. 94, 230504 (2005).
\bibitem{decoy-Wang} X.-B. Wang, Phys. Rev. Lett. 94, 230503 (2005).

\bibitem{Curty10} M. Curty, X. Ma, B. Qi, and T. Moroder, Phys. Rev. A 81, 022310 (2010).
\bibitem{Pas-Mauerer07} W. Mauerer and C. Silberhorn, Phys. Rev. A 75, 050305(R) (2007).
\bibitem{Pas-Adachi07} Y. Adachi, T. Yamamoto, M. Koashi, and N. Imoto, Phys. Rev. Lett. 99, 180503 (2007).
\bibitem{Pas-Ma08} X. Ma and H.-K. Lo, New J. Phys. 10, 073018 (2008).
\bibitem{Pas-Curty09} M. Curty, T. Moroder, X. Ma, and N. L\"{u}kenhaus, Opt. Lett. 34, 3238 (2009).
\bibitem{Tan11} Y. G. Tan and Q. Y. Cai, Int. J. Quantum Inf. 9, 903 (2011).
\bibitem{Renner07} R. Renner, Nature Physics 3, 645 (2007).

\bibitem{Serfling74} R. J. Serfling, The Annals of Statistics, 2(1):39¨C48, (1974).
\bibitem{Dupuis14} F. Dupuis, M. Berta, J. Wullschleger, and R. Renner, Commun. Math. Phys. 328, 251 (2014).
\bibitem{SPDC00} N. L\"{u}tkenhaus, Phys. Rev. A 61, 052304 (2000).
\bibitem{decoy-Ma} X. F. Ma, B. Qi, Y. Zhao, and H. K. Lo, Phys. Rev. A 72, 012326 (2005).
\bibitem{Wang10} J.-Z. Hu and X.-B. Wang, Phys. Rev. A 82, 012331 (2010).
\bibitem{Fung10} C.-H. F. Fung, X. Ma and H. F. Chau, Phys. Rev. A 81, 012318 (2010).
\bibitem{Krapick14}S. Krapick, M.S. Stefszky, M. Jachura, B. Brecht, M. Avenhaus, and C. Silberhorn, Phy. Rev. A 89, 012329 (2014).

\end{thebibliography}
\end{document}